# X-ray outburst of the peculiar Seyfert galaxy IC 3599[*]

**D. Grupe**[1, **], **K. Beuermann**[1,2], **K. Mannheim**[1], **N. Bade**[3, ***], **H.-C. Thomas**[4], **D. de Martino**[5], and **A. Schwope**[6,***]

[1] Universitäts-Sternwarte, Geismarlandstr. 11, D-37083 Göttingen, FRG, E-Mail: GRUPE@USW052.DNET.GWDG.DE
[2] MPI für Extraterrestrische Physik, Giessenbachstr. 6, D-80740 Garching, FRG
[3] Hamburger Sternwarte, Gojenbergsweg 112, D-21029 Hamburg, FRG
[4] MPI für Astrophysik, Karl-Schwarzschild-Str. 1, D-85740 Garching, FRG
[5] ESA/VILSPA, PO Box 50727, E-28080 Madrid, Spain
[6] Astrophysikalisches Institut Potsdam, An der Sternwarte 16, D-14482 Potsdam, FRG

**Abstract.** We report optical, soft X-ray, and UV observations of the peculiar Seyfert galaxy IC 3599 using data obtained with ROSAT and IUE. Most remarkably, we discovered a rapid decrease of the X-ray flux by a factor of about 100 within one year and a more gradual decrease thereafter. The X-ray spectrum of IC3599 was soft at flux maximum and became even softer as the flux decreased. Simultaneously with the late decrease of the X-ray flux, we observed a decrease in the strength of highly ionized optical iron lines. We discuss several explanations for this behaviour including an accretion disk instability and tidal disruption of a star orbiting a central massive black hole.

## 1. Introduction

The ROSAT All-Sky Survey (RASS, Voges 1992) has led to the discovery of numerous AGN with very soft X-ray spectra (Thomas et al. 1992, Walter and Fink 1993). While some of the sources are well-known from previous X-ray experiments and from their appearance in other wavelength bands, others are optically faint and inconspicuous. The fact that virtually unkown galaxies are among the bright soft RASS sources and proved to be faint in follow-up ROSAT observations indicates the importance of large-amplitude X-ray variability in some AGN. In fact, variability over a wide range of time scales is common in AGN (e.g., Miller & Wiita 1991), but amplitudes reaching a factor of 100 as reported here have not been observed before the RASS. Yaqoob et al. (1994) discovered a change in the soft X-ray luminosity of PG 1211+143 by a factor of $\sim 16$ and could rule out variable cold or warm absorbers as the origin of the variability. Most probably, the variability of that source is intrinsic to the central source. The latter could be an accretion disk surrounding a massive black hole (Shakura & Sunyaev 1973), but the most simple $\alpha$-disk models fail if confronted with the results of recent multi-wavelength monitoring of selected AGN (e.g., Clavel et al. 1991, Walter & Courvoisier 1990). More recent models of accretion disks taking into account X-ray irradiation can possibly alleviate the problem (e.g., Ross & Fabian 1993, Mannheim 1995).

IC 3599 = Zw159-34 is a nearby active galaxy ($z = 0.021$, $V = 15.7$) in the outskirts of the Coma cluster of galaxies (Tifft & Gregory 1973, Kent & Gunn 1982) with optical position as given in Table 1. The host galaxy has an extent of $\sim 20''$ diameter and was classified as S0pec by Tifft & Gregory. IC3599 was discovered in the RASS with the X-ray telescope as RX J1237.6+2542, with the wide-field camera as RE J1237+264 (Pounds et al. 1993), and in the optical Hamburg Survey (Bade et al. 1995). IC3599 was not previously seen as an X-ray source; in particular, it was not detected in the HEAO-1 A-2 soft X-ray survey (Nugent at al. 1983) with a limiting sensitivity at this position of $\sim 3\,10^{-14}$ W m$^{-2}$.

**Table 1.** Optical and X-ray positions of IC3599 = Zw159-34.

| Range | $\alpha_{2000}$ | $\delta_{2000}$ |
| --- | --- | --- |
| Optical | $12^h37^m41^s\!.2$ | $26°\ 42'\ 27''$ |
| ROSAT survey | $12^h37^m40^s\!.9$ | $26°\ 42'\ 24''$ |
| ROSAT pointed 1992 | $12^h37^m40^s\!.7$ | $26°\ 42'\ 25''$ |

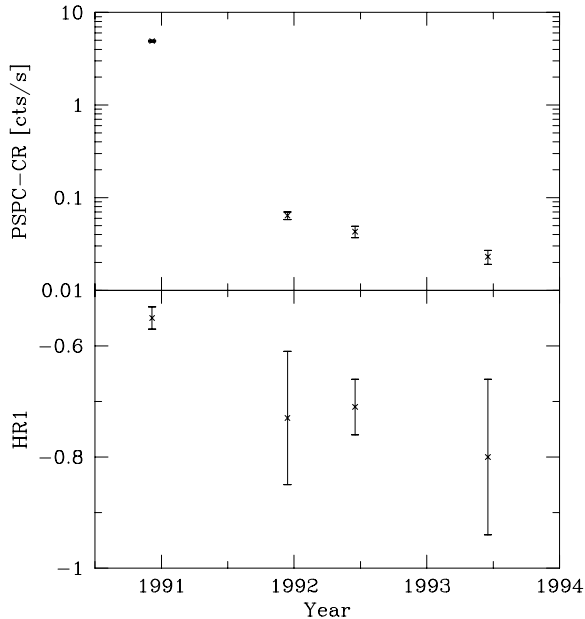

**Fig. 1.** Temporal evolution of the countrate and hardness ratio of IC 3599. The first data point is from the RASS, the remaining ones from pointed ROSAT observations.

## 2. Observations and results

### 2.1. ROSAT–PSPC observations

*Countrates and Hardness ratios:* IC 3599 was one of the brightest AGN in the RASS. During the 2-day RASS coverage on 10/11 December 1990, it was detected with a mean countrate of 3.6 cts s$^{-1}$ with no significant variability. With the vignetting correction, the on-axis countrate becomes $4.9 \pm 0.1$ cts s$^{-1}$. IC3599 was unusually soft for an AGN with a hardness ratio HR = $\frac{\text{hard}-\text{soft}}{\text{hard}+\text{soft}} = -0.55 \pm 0.02$, with soft and hard countrates in the energy bands 0.1 - 0.4 keV and 0.4 - 2.4 keV, respectively.

In order to reduce the errors in the X-ray position derived from the RASS observation, we re-analysed the survey data by using only the photons with energies larger than 0.4 keV and selecting only times when the source appeared in the central part of the detector (15 arcmin radius) where the blurring due to the imaging errors of the telescope are minimal. Our experience is that the survey positions for bright sources are typically better than 10″. The excellent positional agreement with IC3599 leaves no doubt that the intense X-ray emission was associated with this galaxy.

In addition to the survey data and our own pointed observations of June 1992 and June 1993, we used a short pointed observation of December 1991 retrieved from the ROSAT Public Archive. The temporal evolution of the countrate and hardness ratio are shown in Fig. 1. The countrate decreases by a factor of $\sim 80$ from the RASS observation to that of December 1991 with 0.064 cts s$^{-1}$. The decrease continues during June 1992 (0.043 cts s$^{-1}$) and June 1993 (0.023 cts s$^{-1}$) and a correlated decrease of the hardness ratio is observed. Since the flux observed during the RASS exceeds the limit from the HEAO-1 survey we interpret the observed decline as the later part of an outburst.

*PSPC spectra:* Spectral fits for the energy range 0.1 - 2.4 keV were performed for a variety of spectral models. We used a power-law with energy spectral index $\alpha_E$, a blackbody, and a combination of both, absorbed by a column density $N_H$ of cold material of solar abundance. $N_H$ was fixed at the galactic value $N_{H,gal} = 1.3\,10^{20}$ cm$^{-2}$ (Dickey & Lockman 1990). The blackbody is taken to represent the Wien part of any optically thick thermal emitter, although we realize that this part of the spectrum may deviate from a blackbody. The ROSAT PSPC, however, is not sensitive to subtle spectral differences. Table 2 summarizes the results of the spectral fitting. The energy fluxes in the ROSAT band are seen to be rather independent of the spectral model. With $z = 0.021$ and $H_o = 75$ km s$^{-1}$, the implied X-ray luminosity during the RASS is $5\,10^{36}$ W. It decreases to $\sim 6\,10^{34}$ W in 1991 and $\sim 3\,10^{34}$ W in 1993. The bolometric corrections are uncertain, but would be small for a blackbody-like spectrum.

### 2.2. Optical, UV and IR observations

IC 3599 was observed in February 1992 with the 2.2m telescope at Calar Alto/Spain, in February 1993 with the 3.5m telescope at Calar Alto, and in March 1994 and March 1995 with the 2.1m telescope at the McDonald Observatory/Texas with spectral resolutions of ∼8, 8, 4, and 9 Å FWHM, respectively, and with slit widths between 1.1 and 2.0″ (Fig. 2). All observations show a spectrum with narrow Balmer and Helium lines, as well as forbidden lines of oxygen, nitrogen, sulphor, and iron (Tab. 3). The underlying continuum is mostly that of the host galaxy with Ca H+K and Na D in absorption. The FWHM of the Balmer and [OIII] lines corrected for the instrumental resolution are 600 and 200 km s$^{-1}$ respectively, in 1992 and 300 and 200 km s$^{-1}$ in 1995, with errors not exceeding 100 km s$^{-1}$. The widths of the well-detected iron lines are ∼500 km s$^{-1}$. In 1995, the diagnostic line ratios, log([OIII]$\lambda$5007/H$\beta$) $\simeq$ 0.53 and log([NII]$\lambda$6583/H$\alpha$) $\simeq$ −0.50, place IC3599 close to the border-line between Seyfert 2 and HII-galaxies in the diagnostic diagrams of Veilleux and Osterbrock (1987). The X-ray outburst and the presence of the highly ionized iron lines argue for the presence of an active nucleus. We prefer, therefore, a Seyfert 2 classification. In 1992, fairly strong [OI]$\lambda\lambda$6300, 6364, [FeVII]$\lambda\lambda$5721,6087 and [FeX]$\lambda$6375 emission lines were present, which are of lower intensity or even absent in the later years. The rest wavelength of the well-defined 1992 [OI]+[FeX] feature was 6375 Å while that of the very weak 1995 feature was 6369 Å, indicating a decrease of the [FeX] flux by a factor of $\gtrsim$ 5 (Tab. 3). [OIII]$\lambda$4363 seems to be detected and to be unusually strong, a result that requires a separate discussion.

**Table 2.** Parameters of the spectral fits to the X-ray spectra of IC3599 using $N_H = N_{H,gal}$. The unabsorbed flux $F_E$ in the ROSAT-PSPC band (0.1-2.4 keV) is given in units of $10^{-16}$ W m$^{-2}$ and the blackbody temperature kT in eV.

|  | RASS | 1991.9 | 1992.5 | 1993.5 |
|---|---|---|---|---|
| *Power law* | | | | |
| $\alpha_E$ | 2.1 | 3.3 | 2.3 | 3.1 |
| $F_{E,pl}$ | 631 | 12.6 | 8.5 | 4.3 |
| $\chi^2/\nu$ | 135/43 | 25/11 | 15/13 | 6/5 |
| *Blackbody* | | | | |
| $kT$ | 94 | 65 | 80 | 62 |
| $F_{E,bb}$ | 530 | 6.3 | 5.8 | 2.2 |
| $\chi^2/\nu$ | 75/43 | 15/11 | 30/13 | 4/5 |
| *Power law plus blackbody* | | | | |
| $\alpha_E$ | 1.0 (fix) | – | 1.0 (fix) | – |
| $F_{E,pl}$ | 55 | – | 1.6 | – |
| $kT$ | 91 | – | 65 | – |
| $F_{E,bb}$ | 480 | – | 4.5 | – |
| $F_{E,sum}$ | 535 | – | 6.1 | – |
| $\chi^2/\nu$ | 59/42 | – | 12/12 | – |

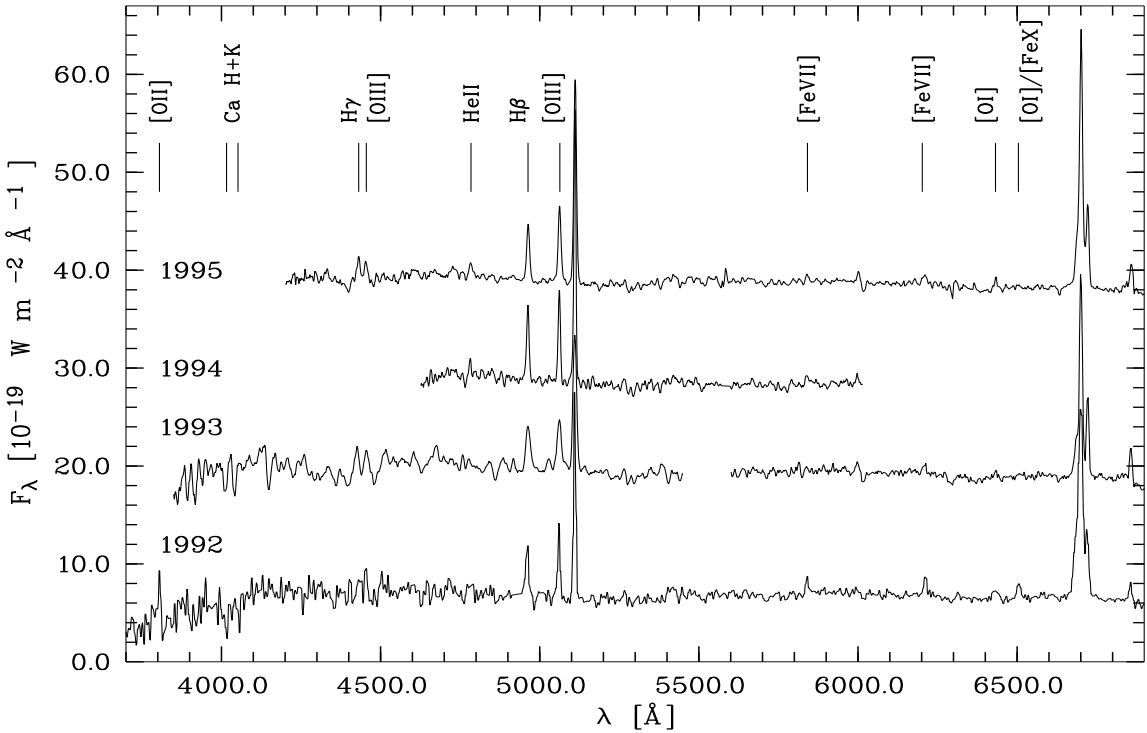

**Fig. 2.** Optical spectra of IC 3599. The spectra are plotted with offsets 10, 20, and 30 with respect to the 1992 spectrum.

A 170 min exposure of IC 3599 was obtained with the IUE-satellite simultaneously with ROSAT on June 18, 1992. The resulting spectrum shows only a weak continuum and emission of L$\alpha$. The continuum flux $F_\lambda$ is roughly constant from the optical to the UV. With the IRAS-satellite only weak emission at 60 $\mu$m was detected. The ADDSCAN procedure of the IPAC yields $F_{60\mu} = 0.11$ Jy.

**Table 3.** Intensities of emission lines in the 1992 − 1995 spectra of IC3599. Values are given in units of $2.2\,10^{-19}$ W m$^{-2}$. [1]

|  | 1992.1 | 1993.1 | 1994.2 | 1995.2 |
|---|---|---|---|---|
| HeII$\lambda$4686 | 16 ± 8 | 8 ± 8 | 12 ± 6 | 16 ± 3 |
| H$\beta$ | 35 ± 8 | 35 ± 8 | 33 ± 6 | 29 ± 3 |
| [OIII]$\lambda$4959 | 36 ± 8 | 37 ± 8 | 37 ± 6 | 40 ± 3 |
| [OIII]$\lambda$5007 | 100 | 100 | 100 | 100 |
| [FeVII]$\lambda$5721 | 10 ± 2 | 5 ± 2 | 5 ± 2 | 3 ± 1 |
| [FeVII]$\lambda$6087 | 12 ± 2 | 6 ± 2 | — | 7 ± 2 |
| [OI]$\lambda$6300 | 11 ± 3 | 4 ± 2 | — | 5 ± 2 |
| [OI]$\lambda$6364 / [FeX]$\lambda$6375 | 13 ± 2 | 3 ± 2 | — | 2 ± 1 |
| H$\alpha$ | 219 ± 23 | 135 ± 16 | — | 152 ± 18 |
| [NII]$\lambda$6584 | 46 ± 11 | 45 ± 13 | — | 48 ± 9 |

[1] Observed 1992 and 1993 intensities were multiplied by factors of 1.22 and 0.86, respectively, to adjust to an adopted constant [OIII]$\lambda$5007 flux.

## 3. Discussion

IC 3599 was observed in the optical by Tifft and Gregory (1973). Their spectrum shows strong [OII]$\lambda$3727, narrow H$\beta$, and [OIII]$\lambda\lambda$4959, 5007 emission lines, which are also found in the spectra obtained by us some 20 years later. Following the X-ray outburst observed during 1991 no convincing variation of the optical continuum and [OIII] line flux was found during our follow-up observations between February 1992 and March 1995. However, we discovered a significant decrease of the flux of [FeX]$\lambda$6375 and of [FeVII]$\lambda\lambda$5721,6086 by factors of $\gtrsim 5$ and $\sim 2$, respectively, along with a monotonic fading of the X-ray flux. In addition, the Balmer line flux may be decreasing. These findings suggest that the emission lines are due to photoionization by a varying continuum. The characteristic radius of the high-ionization iron line region implied by the light travel time is of the order of $R \sim 1$ lyr or less.

The interpretation of the high X-ray flux during the RASS as an outburst rather than a short-term disappearance of an absorbing screen is strengthened by the over-all energetics, since the luminosities radiated in the far-infrared, optical and UV ranges are much less than the RASS luminosity. Reprocessing of the X-rays by the absorbing gas would inevitably lead to re-radiation in either of these frequency ranges. Thus, if the thermal energy release corresponds to accretion onto a massive black hole, the accretion rate must have undergone dramatic changes to cause the outburst.

We can bracket the mass $M$ of the putative black hole assuming that (i) the X-ray luminosity observed during the RASS $L_{\rm RASS} \sim 5\,10^{36}$ W does not exceed the Eddington luminosity and (ii) the observed iron line width of $\sim$500 km s$^{-1}$ corresponds to an average velocity greater than or equal to the Keplerian velocity in the gravitational field of the black hole, i.e. FWHM/$2\sqrt{\ln 2} \geq \sqrt{GM/R}$. This yields a mass range of $0.4 \lesssim m_6 \lesssim 10\,(R/1\,{\rm lyr})$ where $m_6 = M/10^6 {\rm M}_\odot$. A rather low mass $m_6 \sim 1$ would be in agreement with the fact that the X-ray observations are best explained by thermal spectra with $kT \sim$ 60-100 eV.

One possibility to obtain a higher accretion rate for a short period of time is a *thermal (or other) instability* in an accretion disk (e.g., Honma et al. 1991). Disk instabilities have been proposed as the origin of X-ray outbursts in accreting binary stars but also to explain the optical/UV variations in AGN (Siemiginowska & Czerny 1989). In its normal state, the disk has a rather low viscosity until the instability is switched on at a radius $R_{\rm d}$ and the disk attains a high-viscosity state. In this state, matter is rapidly accreted leading to a high-luminosity outburst which decays over the viscous time scale. Adopting $m_6 = 1$, $\alpha = 1$, $L_{\rm bol} \simeq L_{\rm x,quiet} \simeq 3\,10^{34}$ W and $r = R_{\rm d}/R_{\rm G} = 10$ where $R_{\rm G} = 1.5\,10^9 m_6$ m, the viscous time scale $t_{\rm visc} \propto \alpha^{-4/5} L_{\rm bol}^{-3/10} m_6^{3/2} r^{5/4}$ is of the order of $\sim 1$ year (Frank et al. 1985). Thus, a very low value for the black hole mass would be required by this mechanism.

Another possibility to feed the central engine at a high rate for a short period of time is the *tidal disruption of a star* orbiting near a supermassive black hole. The estimated timescale for a single event where the debris of a star is swallowed by a $10^6$ M$_\odot$ black hole is of the order of $\sim 1$ year (Rees 1988). The fading countrates during our pointed X-ray observations and the decrease in iron line flux would indicate that the mass of the disrupted star is used up, so that accretion settles to a lower state.

Finally, we note that a single supernova outburst can not account for the observed peak X-ray flux. Assuming that the X-ray emission arises from the interaction of the ejected matter with the dense (red-giant) wind, the X-ray luminosity scales as $L_{\rm x} \propto \dot{M}_{\rm wind}\, v_{\rm wind}^{-1} T_{\rm x}^{3/2}$ (e.g. Bregman & Pildis 1992). Given the softness of the observed X-ray emission from IC3599, the X-ray luminosity is limited to $L_{\rm x} < 10^{33}$ W. Clumpiness may somewhat raise this limit.

Without detailed modelling, further conclusions are not possible at this stage. Nevertheless, it is clear that the ROSAT survey with its snapshot views of the sky has given us the opportunity to witness a very energetic moment in the life of an AGN, providing crucial information about the way in which the central engine if fed.

*Acknowledgements.* We thank Wolfram Kollatschny for useful comments and Boris Gänsicke for help with the data reduction. This research has made use of the NASA/IPAC Extragalactic Database (NED) and of the Infrared Processing and Analysis Center at Caltech (IPAC). This research was supported by the DARA under grant 50 OR 92 10.

*Note added in proof:* After submission of this paper, we received a preprint by Brandt et al. who performed a similar study starting from the detection of IC3599 with the ROSAT-WFC. Their optical spectrum is complementary to ours, being taken in May 1991, somewhat earlier in the decline of the X-ray outburst. This spectrum shows Balmer and forbidden iron lines roughly an order of magnitude more intense than in our spectra, thus confirming our conclusion of optical line variability of IC3599 as a consequence of the X-ray outburst.